\documentclass[aps,prl,amsfonts,amssymb,twocolumn,amsmath,preprintnumbers,nofootinbib,floatfix,superscriptaddress]{revtex4-2}%
\usepackage{amsfonts}
\usepackage{mathrsfs}
\usepackage{amsmath}
\usepackage{color}
\usepackage{graphicx}
\usepackage{bm}
\usepackage{amssymb}
\usepackage{xspace}
\usepackage{epstopdf}
\usepackage{dcolumn}
\usepackage{longtable}
\usepackage{multirow}
\usepackage{float}
\usepackage{comment}

\usepackage[colorlinks=true, letterpaper=true, pdfstartview=FitV, linkcolor=blue, citecolor=blue, urlcolor=blue]{hyperref}

\makeatother
\begin{document}

\title{Intrinsic Nonlinear Planar Hall Effect}

\author{Yue-Xin Huang}
\thanks{These authors contributed equally to this work.}
\affiliation{Research Laboratory for Quantum Materials, Singapore University of Technology and Design, Singapore 487372, Singapore}

\author{Xiaolong Feng}
\thanks{These authors contributed equally to this work.}
\affiliation{Research Laboratory for Quantum Materials, Singapore University of Technology and Design, Singapore 487372, Singapore}

\author{Hui Wang}
\affiliation{Research Laboratory for Quantum Materials, Singapore University of Technology and Design, Singapore 487372, Singapore}

\author{Cong Xiao}
\email{congxiao@hku.hk}
\affiliation{Department of Physics, The University of Hong Kong, Hong Kong, China}
\affiliation{HKU-UCAS Joint Institute of Theoretical and Computational Physics at Hong Kong, China}

\author{Shengyuan A. Yang}
\affiliation{Research Laboratory for Quantum Materials, Singapore University of Technology and Design, Singapore 487372, Singapore}

\begin{abstract}
We propose an intrinsic nonlinear planar Hall effect, which is of band geometric origin, independent of scattering, and scales
with the second order of electric field and first order of magnetic field.
We show that this effect is less symmetry constrained compared to other nonlinear transport effects and is supported in a large class of nonmagnetic polar and chiral crystals. Its characteristic
angular dependence provides an effective way to control the nonlinear output. Combined with first-principles calculations, we evaluate this
effect in the Janus monolayer MoSSe and report experimentally measurable results.
Our work reveals an intrinsic transport effect, which offers a new tool for material characterization and a new mechanism for nonlinear device application.
\end{abstract}
\maketitle

\emph{Intrinsic} transport properties, which are independent of scattering and solely manifest the band structure geometry, have been a focus in condensed matter physics research. As prominent examples, the intrinsic contributions to the anomalous Hall effect~\cite{Jungwirth2002,Nagaosa2002,Nagaosa2010}, spin Hall effect~\cite{murakami_dissipationless_2003,sinova_universal_2004,Sinova2015}, and anomalous Nernst effect~\cite{xiao_berry-phase_2006} are connected to the Berry curvature of the bands~\cite{Xiao2010} and found great success in explaining experimental observations.

In this work, we reveal a new intrinsic transport in the \emph{nonlinear} planar Hall effect (NPHE). The planar Hall effect (PHE) refers to the setup where the applied magnetic $B$ field is inside the transport plane formed by the transverse response current $j_\text{H}$ and the driving $E$ field. This setup dictates that the response is distinct from the ordinary Hall effect, as the Lorentz force is not in action here.
The linear PHE was initially studied in magnetic materials~\cite{tang_giant_2003,bowen_order-driven_2005,seemann_origin_2011}, as a useful tool for probing the magnetization reversal. Later, it received interest in nonmagnetic materials, especially topological materials such as Weyl semimetals~\cite{Nandy2017,kumar_planar_2018,deng_quantum_2019,ma_planar_2019} and topological insulators~\cite{breunig_gigantic_2017,wu_oscillating_2018,rakhmilevich_unconventional_2018}. More recently, NPHE, the nonlinear version of the effect with $j_\text{H}\sim E^2 B$, was reported in experiment~\cite{He2019,esin_nonlinear_2021}. Notably, \citet{He2019} observed the NPHE at the two-dimensional (2D) surface of a topological insulator and connected it to the spin-momentum locking of the topological surface states. A few theories were put forward to explain the effect~\cite{He2019,zheng_origin_2020,rao_theory_2021,battilomo_anomalous_2021}. However, it is important to note that all the NPHE mechanisms proposed  so far are of \emph{extrinsic} nature, i.e., they depend on scattering, which can be readily seen as their contributions to the the Hall current scale as $\sim \tau^2$ in the electron relaxation time $\tau$.

Here, we propose the \emph{intrinsic} NPHE, which is completely determined by the band structure therefore represents an intrinsic material property. We show that the intrinsic NPHE tensor is expressed in terms of the band geometric quantities including the Berry-connection polarizability (BCP)~\cite{Gao2014,Liu2022} and its spin susceptibility. We clarify the symmetry property of the effect and find that unlike the linear PHE or nonlinear anomalous Hall effect which are strongly constrained by symmetry, the intrinsic NPHE can exist in a wide range of non-centrosymmetric crystals. Moreover, the response exhibits a characteristic angular dependence in the applied fields, providing a convenient switch for the nonlinear output signal in applications. As a unique advantage, intrinsic effects allow a quantitative evaluation. We combine our theory with first-principles calculations and demonstrate a sizable intrinsic NPHE in a famous polar 2D material, the monolayer MoSSe. Our work discovers a new intrinsic transport property, which offers a tool for characterizing fundamental band geometry of materials and holds great potential for nonlinear device applications.

\textcolor{blue}{\textit{Origin of intrinsic NPHE.}} Consider a 2D nonmagnetic system occupying the $x$-$y$ plane. First, note that since the current ($E$ field) is odd (even) under time reversal operation, the intrinsic transport cannot occur without the applied $B$ field. Second, in the 2D planar Hall setup, the $B$ field couples only to the electron spin via Zeeman coupling but not to its orbital motion. Hence, the effect of $B$ field here can be captured as a perturbation to the band structure, which causes a spin splitting and transforms the original nonmagnetic band structure to a kind of ``magnetic'' band structure.

Starting from this perturbed band structure, the intrinsic NPHE of the original system is equivalent to an
intrinsic nonlinear anomalous Hall effect of the new (perturbed) system. It
can be most readily derived in the framework of semiclassical theory~\cite{sundaram_wave-packet_1999,Xiao2010}. Like the intrinsic anomalous Hall effect, in this framework, it is connected to the so-called anomalous velocity of electrons $\bm v^A=\bm E\times\bm{\Omega}$ (set $e=\hbar=1$)~\cite{karplus_hall_1954,chang_berry_1995}, which is evidently transverse to the $E$ field. Here, $\bm{\Omega}$ is the Berry curvature, which points out of plane for a 2D system. Note that to study the nonlinear response at the $E^2$ order, one has to include in $\bm{\Omega}$ the field correction of the Berry curvature $\bm{\Omega}^E$ that is linear in $E$. Such a field correction has been captured in the recently developed extended semiclassical theory~\cite{Gao2014,Gao2015}. The result shows that $\bm{\Omega}^E=\nabla_{\bm k}\times\boldsymbol{\mathcal{A}}^{E}$, where ${\mathcal{A}}^{E}_a=\tilde{G}_{ab}E_{b}$ is the field corrected Berry connection, $\tilde{G}_{ab}$ is a gauge-invariant quantity known as the BCP tensor, the indices $a, b$ label the Cartesian components and repeated indices are summed over.

The intrinsic nonlinear Hall current is then obtained as $\bm j^\text{int}=\int f_0(\tilde{\varepsilon})\bm E\times \bm{\Omega}^E$, with $f_0$ the {equilibrium} Fermi distribution function. In this result, the material's information is encoded in the band energy $\tilde{\varepsilon}$ and the BCP $\tilde{G}_{ab}$, but note that here they are defined with respect to the $B$-perturbed band structure (which is why we add a tilde in these symbols). By straightforward calculations, we can expand them to the $O(B^1)$ order and express all quantities using the original \emph{unperturbed} eigenstates $|u_n(\bm k)\rangle$ and band energies $\varepsilon_n(\bm k)$. For instance, we have
\begin{equation}
\tilde{G}_{ab}=G_{ab}+\Lambda_{abc}B_{c},
\end{equation}
where $G_{ab}$ is the BCP for the unperturbed system, and $\Lambda_{abc}=\partial_{B_{c}%
}\tilde{G}_{ab}|_{B=0}$ can be interpreted as the spin susceptibility of BCP. Explicitly, for a band with index $n$,
\begin{equation}
G_{ab}^{n}(\bm k)=2\mathrm{{\operatorname{Re}}}\sum_{m\neq n} \frac{v_{a}^{nm}%
v_{b}^{mn}}{(\varepsilon_{n}-\varepsilon_{m})^{3}},
\end{equation}
and
\begin{widetext}
\begin{align}
\Lambda_{abc}^{n}(\bm k) =2\mathrm{{\operatorname{Re}}}\sum_{m\neq n}\bigg[\frac
{3v_{a}^{nm}v_{b}^{mn}\left(  \mathcal{M}_{c}^{n}-\mathcal{M}_{c}^{m}\right)
}{(\varepsilon_{n}-\varepsilon_{m})^{4}}
-\sum_{\ell\neq n} \frac{\left(
v_{a}^{\ell m}v_{b}^{mn}+v_{b}^{\ell m}v_{a}^{mn}\right)  \mathcal{M}%
_{c}^{n\ell}}{\left(  \varepsilon_{n}-\varepsilon_{\ell}\right)  (\varepsilon
_{n}-\varepsilon_{m})^{3}}
-\sum_{\ell\neq m} \frac{\left(
v_{a}^{\ell n}v_{b}^{nm}+v_{b}^{\ell n}v_{a}^{nm}\right)  \mathcal{M}%
_{c}^{m\ell}}{\left(  \varepsilon_{m}-\varepsilon_{\ell}\right)  (\varepsilon
_{n}-\varepsilon_{m})^{3}}\bigg],
\end{align}
\end{widetext}
where  $v_{a}^{nm}=\langle u_{n}|\hat{v}_{a}\mathbf{|}u_{m}\rangle$ is the
velocity matrix element,
$\boldsymbol{\mathcal{M}}^{mn}=-g\mu_{B}\boldsymbol{s}^{mn}$ is the spin magnetic moment matrix element, with $g$ the $g$-factor, $\mu_B$ the Bohr magneton, and $\boldsymbol{s}^{mn}$
the spin matrix element; and $\boldsymbol{\mathcal{M}}^{n}\equiv \boldsymbol{\mathcal{M}}^{nn}$ denotes the diagonal element. Meanwhile, $\tilde{\varepsilon}_n(\bm k)={\varepsilon}_n-\boldsymbol{\mathcal{M}}^{n}\cdot\boldsymbol{B}$. Substituting these expansions into the nonlinear Hall current and keeping terms up to $O(E^2B)$, one obtains the intrinsic NPHE current $\bm j^\text{int}$ (details in the Supplemental Material~\cite{supp}).

The result is cast into an intrinsic NPHE conductivity $\chi^{\text{int}}$ defined by
\begin{equation}\label{jj}
j_{a}^{\text{int}}=\chi_{abcd}^{\text{int}}E_{b}E_{c}B_{d},
\end{equation}
where all the indices $\in\{x,y\}$. Then, we  find
\begin{equation}
\chi_{abcd}^{\text{int}}=\sum_{n}\int\frac{d^{2}\boldsymbol{k}}{\left(
2\pi\right)  ^{2}}f_{0}^{\prime}\ \alpha_{abcd}^{n}\left(  \boldsymbol{k}%
\right)  , \label{central}%
\end{equation}
with
\begin{equation}\label{alpha}
\alpha_{abcd}^{n}=(v_{a}^{n}\Lambda_{bcd}^{n}-v_{b}^{n}\Lambda_{acd}%
^{n})+\left(  \partial_{a}G_{bc}^{n}-\partial_{b}G_{ac}^{n}\right)
\mathcal{M}_{d}^{n}.
\end{equation}
Evidently, $\chi^{\text{int}}$ is antisymmetric in its first two indices, which ensures
$j_{a}^{\text{int}}E_{a}=0$, so $j_{a}^{\text{int}}$ is truly a dissipationless
Hall current. The expressions in (\ref{central}, \ref{alpha}) confirm that the effect is a Fermi surface property
determined completely by the material's band structure, manifesting its intrinsic nature. In Eq.~(\ref{alpha}), the first term of $\alpha$
is an antisymmetrized product of the band velocity $v_{a}^{n}$ and the spin susceptibility of
BCP; whereas the second term represents a product of the \textit{k}-space dipole of BCP and the spin magnetic moment.

As emphasized, the intrinsic NPHE $\chi^{\text{int}}$ is independent of scattering. This is distinct from previously
discussed extrinsic contributions which explicitly involve scattering and scale as $\tau^2$~\cite{He2019,rao_theory_2021,battilomo_anomalous_2021} (note that contribution $\sim \tau E^2B$ is forbidden for nonmagnetic systems by the time reversal symmetry). In practice, the different contributions can be readily separated by their different $\tau$ scaling.

\textcolor{blue}{\textit{Symmetry property and angular dependence.}} Crystal symmetries constrain the form of the $\chi^{\text{int}}$ tensor. Clearly, $\chi^{\text{int}}$ would take its simplest form in a coordinate system adapted to the crystal axis~\cite{Kang2019,Lai2021}. For example, for a rectangular lattice, one would prefer to choose $x$ and $y$ along the $a$ and $b$ axis for the calculation of $\chi^{\text{int}}$. In real experiments, the applied $E$ and $B$ fields are in general not aligned with the crystal axis, and one typically rotates these fields in the plane.
To analyze such general cases, we specify $E$
($B$) field by its polar angle $\theta$ ($\varphi$) from the $x$
direction (which is already chosen as some crystal axis), i.e.,
$(E_x, E_y)=E\left(  \cos\theta,\sin\theta\right)  $ and $(B_x,B_y)
=B\left(  \cos\varphi,\sin\varphi\right)$, then the intrinsic NPHE current is in the direction $\hat{z}\times \bm E$ with a magnitude
\begin{equation}
j^{\text{int}}=\chi^\text{int}_\text{H}(\theta,\varphi)E^{2}B,
\end{equation}
where the angle-dependent scalar coefficient $\chi^\text{int}_\text{H}$ is a combination of the $\chi^{\text{int}}$ tensor elements:
\begin{align}
\chi^\text{int}_{\text{H}} (\theta,\varphi)  =&(\chi_{yxxx}^{\text{int}}\cos\varphi+\chi_{yxxy}%
^{\text{int}}\sin\varphi)\cos\theta\nonumber\\
&  -(\chi_{xyyx}^{\text{int}}\cos\varphi+\chi_{xyyy}^{\text{int}}\sin\varphi
)\sin\theta. \label{response}%
\end{align}
Note that since $\chi^{\text{int}}_{abcd}$ is enforced to be antisymmetric in $a$ and $b$ indices, there are at most four independent tensor elements, which we choose here as $\chi_{yxxx}^{\text{int}}$, $\chi_{yxxy}^{\text{int}}$,
$\chi_{xyyx}^{\text{int}}$ and $\chi_{xyyy}^{\text{int}}$.

\begin{table}[ptb]
\caption{{}Constraint on the $\chi^{\text{int}}$ tensor from
point group symmetries pertaining to 2D materials. \textquotedblleft%
$\checkmark$\textquotedblright\ (\textquotedblleft$\times$\textquotedblright)
means that the element is symmetry allowed (forbidden). Because $\chi
^{\text{int}}$ is time reversal ($\mathcal{T}$) even, symmetry operations
$\mathcal{R}$ and $\mathcal{RT}$ impose the same constraint.}%
\label{tab: group}
\begin{tabular}
[b]{ccccccc}\hline\hline
& $\mathcal{P}$, $S_{6}$, $\sigma_{z}$ & \ \ $C_{2}^{z}$ \ \  & \ \ $S_{4}%
^{z}$ \  & \ \ $C_{3,4,6}^{z}$\ \ \  & $\ C_{2}^{x},$\ $C_{2}^{y}$\ \  &
\ $\sigma_{x},$ $\sigma_{y}$\ \ \\\hline
&  &  &  &  &  & \\
$\chi_{xyyx}^{\text{int}}$ & $\times$ & $\checkmark$ & $\checkmark$ &
$\checkmark$ & $\checkmark$ & $\times$\\
&  &  &  &  &  & \\
$\chi_{yxxx}^{\text{int}}$\  & $\times$ & $\checkmark$ & $\checkmark$ &
$\checkmark$ & $\times$ & $\checkmark$\\
&  &  &  &  &  & \\
$\chi_{xyyy}^{\text{int}}$\  & $\times$ & $\checkmark$ & $\chi_{yxxx}%
^{\text{int}}$ & $-\chi_{yxxx}^{\text{int}}$ & $\times$ & $\checkmark$\\
&  &  &  &  &  & \\
\ $\chi_{yxxy}^{\text{int}}$\ \  & $\times$ & $\checkmark$ & $-\chi_{xyyx}^{\text{int}}$ &
$\chi_{xyyx}^{\text{int}}$ & $\checkmark$ & $\times$\\\hline\hline
\end{tabular}
\end{table}

Next, we analyze the symmetry property of the $\chi^{\text{int}}$ tensor. There are two quick observations. First, the inversion symmetry must be broken, because the $B$ field is even under inversion whereas the current and the $E$ field are odd. Second, the system must not have a horizontal mirror $\sigma_z$, because it would flip the $B$ field but keep other vectors in (\ref{jj}) unchanged. Then, out of the 18 noncentrosymmetric
point groups pertaining to 2D systems, 16 allow the intrinsic NPHE. These include $S_4$, $D_{2d}$, all polar point groups and chiral point groups, i.e., $C_{n}$, $C_{nv}$
($n=1,2,3,4,6$) and $D_{n}$ ($n=2,3,4,6$)~\cite{note}. This indicates that the intrinsic NPHE can be supported in broad crystal systems. In comparison, the constraint on the linear PHE (with $j_\text{H}\sim EB$) is much more stringent: among these groups, it is only allowed in $C_{1}$, $C_{1v}$, and $C_{2}$ (in-plane polar axis). Similarly, the nonlinear anomalous Hall effect by Berry curvature dipole is suppressed in 2D by any symmetry other than a single mirror line~\cite{Fu2015}.

\begin{figure}[tb]
    \centering
    \includegraphics[width=0.4\textwidth]{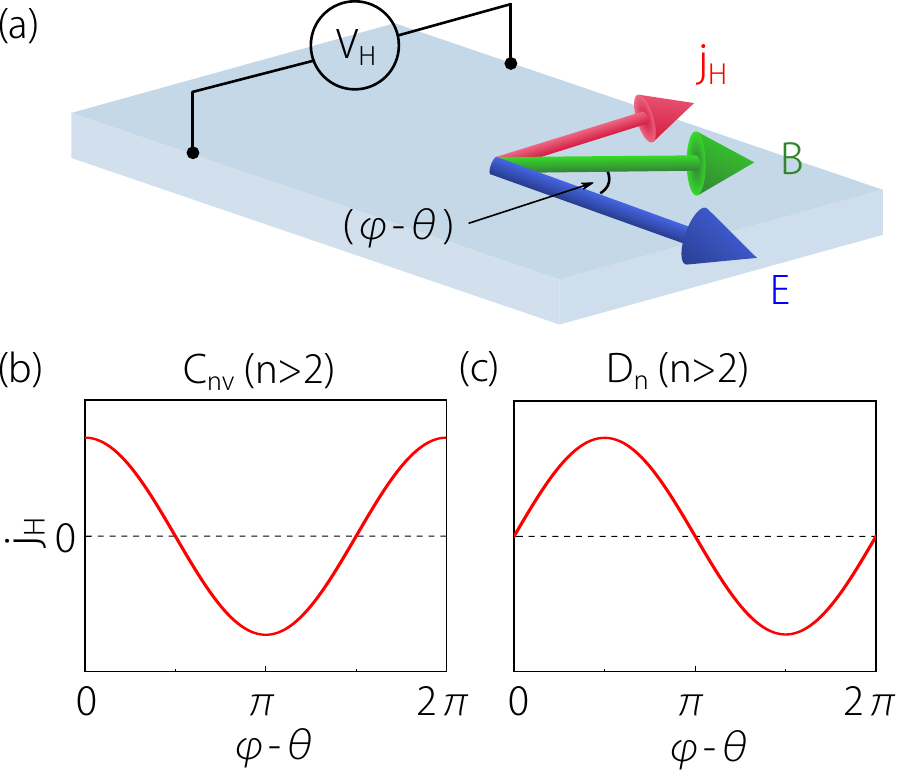}
    \caption{(a) Schematic of the measurement setup for NPHE. (b, c) For a material with
        point group ${C_{nv}}$ or ${D_{n}}$ ($n>2$), the intrinsic NPHE response exhibits a simple cosine or sine
        dependence on the angle $(\varphi-\theta)$ between $E$ and $B$ fields, when $B$ field is rotated in the plane.
        }
    \label{fig-fig1}
\end{figure}

In Table~\ref{tab: group}, we list the constraints on the $\chi^{\text{int}}$ elements by symmetries in the 16 allowed groups. It follows from these results that the intrinsic NPHE response exhibits rich and interesting angular dependence as one rotates the $E$ or $B$ fields. For example, polar
groups $C_{nv}$ with $n>2$ support only one independent tensor element $\chi_{xyyy}^{\text{int}}=-\chi_{yxxx}^{\text{int}}$, and
$\chi_{\text{H}}^\text{int}$ is reduced to
\begin{equation}
\chi_{\text{H}}^\text{int}=\chi_{yxxx}^{\text{int}}\cos(\varphi-\theta), \label{Cnv}%
\end{equation}
which depends only on the relative angle between the ${E}$
and ${B}$ fields through a simple cosine relation.
The response is maximized when the two fields are
parallel or anti-parallel, but vanishes when they are perpendicular. In contrast, materials with dihedral chiral point groups $D_{n}$ with $n>2$ host only $\chi_{xyyx}^{\text{int}}=\chi_{yxxy}^{\text{int}}$, then
\begin{equation}
\chi_{\text{H}}^\text{int}=\chi_{yxxy}^{\text{int}}\sin(\varphi-\theta), \label{Dn}%
\end{equation}
which is maximized when the two fields are perpendicular, but vanishes
when they become collinear. These cases are illustrated in Fig.~\ref{fig-fig1}.

Our above analysis demonstrates the intrinsic NPHE as a promising tool for characterizing a large class of materials not accessible by other nonlinear transport measurements. The clarified angular dependence offers a convenient knob to control the nonlinear response and to switch it on and off by simply rotating the $B$ (or $E$) field (as in Fig.~\ref{fig-fig1}(b,c)). This could be useful for the design of new devices for nonlinear rectification or frequency doubling.

\textcolor{blue}{\textit{A model study.}} To illustrate the features of the
effect and the underlying geometric quantities, we first
apply our theory to a tight-binding model on a honeycomb lattice with $C_{3v}$
symmetry. The model reads
\begin{align}
H=&-t \sum_{\langle ij\rangle,\alpha}c_{i\alpha}^\dagger c_{j\alpha}+\frac{\Delta}{2} \sum_{i,\alpha} \xi_i c^\dagger_{i\alpha} c_{i\alpha}\nonumber\\
  &+i t_R\sum_{\langle ij\rangle,\alpha\beta}(\bm{s}\times {\bm d}_{ij})^z_{\alpha\beta}c^\dagger_{i\alpha}c_{j\beta}, \label{model}
\end{align}
where the first term is the nearest neighbor hopping, the second term is a staggered potential with $\xi_i=\pm 1$ on the two sublattices, the last term is a Rashba spin-orbit coupling (SOC), $\alpha$ and $\beta$ are the spin labels, and ${\bm d}_{ij}$ is a unit vector pointing from site $j$ to $i$. This model is similar to that for silicene or germanene on a substrate (some intrinsic SOC terms
inessential to our discussion are neglected)~\cite{liu_low-energy_2011,liu_quantum_2011}. The staggered potential violates the inversion symmetry and the Rashba SOC breaks the horizontal mirror. They together lower the system symmetry from $D_{6h}$ to $C_{3v}$, which supports the intrinsic NPHE.

\begin{figure}[ptb]
    \centering
    \includegraphics[width=0.45\textwidth]{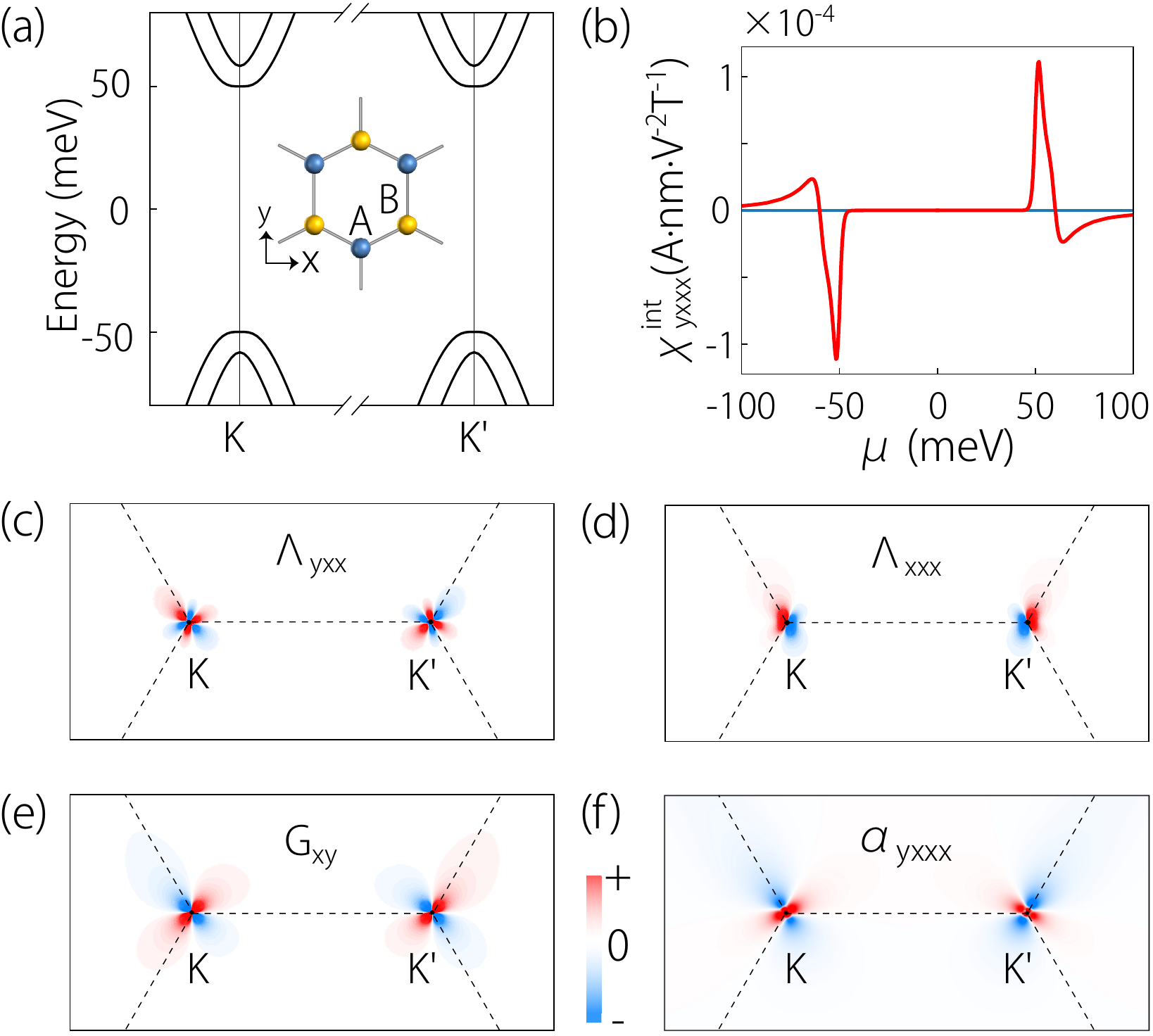}
    \caption{(a) Low-energy bands for the honeycomb lattice model (\ref{model}), with two valleys at $K$ and $K'$. The inset shows the honeycomb lattice, which has two sublattices.
        (b) Calculated response coefficient $\chi_{yxxx}^{\text{int}}$ versus Fermi energy $\mu$.
        (c-f) show the distribution of (c) $\Lambda_{yxx}$, (d) $\Lambda_{xxx}$, (e) $G_{xy}$ and (f) $\alpha_{yxxx}$ for the upper valence band
    in $k$-space. In the calculation, we take $t=1$ eV, $t_R=10$ meV, and $\Delta=100$ meV.}
    \label{fig-fig2}%
\end{figure}

Figure~\ref{fig-fig2}(a) shows the energy spectrum of the model, which exhibits a binary valley structure at $K$ and $K'$. The splitting of the conduction (valence) band is due to the SOC. For the coordinate setup in Fig.~\ref{fig-fig2}(a), only one tensor element $\chi_{yxxx}^{\text{int}}$ is relevant. In Figs.~\ref{fig-fig2}(c)-(f), we plot the $k$-space distribution of relevant band geometric quantities, including $\Lambda_{yxx}$, $\Lambda_{xxx}$, $G_{xy}$ and $\alpha_{yxxx}$, for the upper valence band. One observes that these quantities are concentrated around the two valleys, where the energy difference between the bands is small. This can be understood by noting that similar to the Berry curvature, all these geometric quantities encodes the interband coherence, so they are peaked mainly at small-gap regions. For the model here, the intrinsic NPHE should become large when the Fermi surface crosses the hot spots around the band edge, which is confirmed by the result in Fig.~\ref{fig-fig2}(b).

This example shows that to have a sizable intrinsic NPHE, the system should have its Fermi level near certain band near degeneracies, where the band geometric quantities are pronounced. In addition, a sizable SOC is also a necessary condition. Generally, without SOC, the band structure perturbed by the $B$ field would consist of two decoupled spin channels. Each spin channel can be regarded as a spinless system with an effective time reversal symmetry~\cite{vanderbilt_berry_2018,wu_nodal_2018}, which then suppresses the intrinsic NPHE. This point is also verified in our model calculation.

\begin{figure}[ptb]
    \centering
    \includegraphics[width=0.45\textwidth]{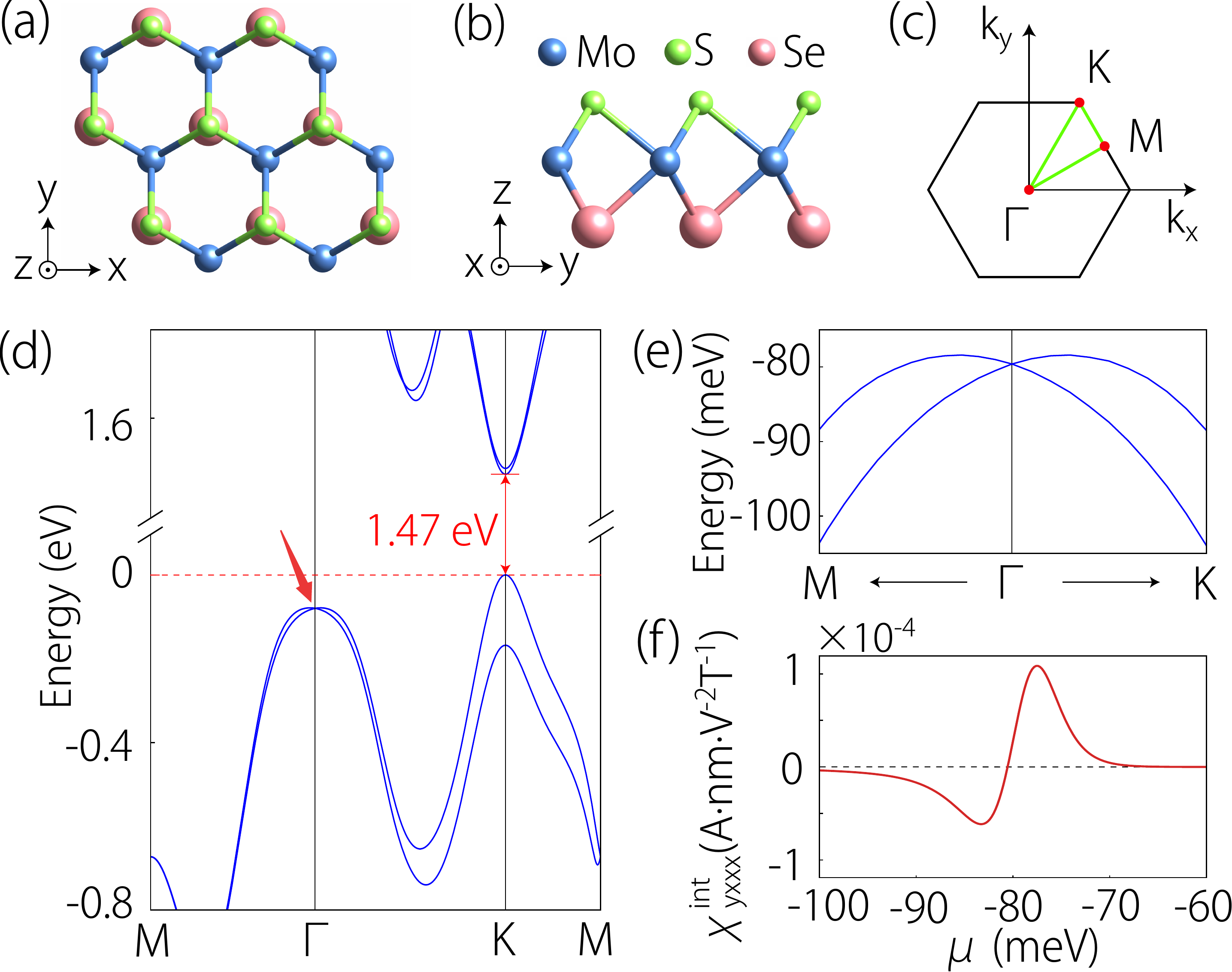}
    \caption{(a) Top and (b) side views of MoSSe monolayer. (c) Brillouin zone of MoSSe.
    (d) Calculated band structure along the high-symmetry paths. (e) Enlarged view around the valence band degeneracy point at $\Gamma$, as indicated by the arrow in (d). (f) Calculated intrinsic NPHE response $\chi_{yxxx}^\mathrm{int}$, which shows a peak for Fermi level around the point in (e).}
    \label{MoSSe}%
\end{figure}

\textcolor{blue}{\textit{Application to 2D MoSSe.}}
As discussed, the advantage of an intrinsic effect is that it can be evaluated from
first principles to yield quantitative predictions in concrete
materials. Here, guided by the symmetry constraint and lessons from the model study,
we investigate the 2D Janus monolayer MoSSe, which is a polar material
synthesized in 2017~\cite{zhang2017janus,lu2017Janus}
and under active research since then~\cite{Dong2017,Guo2021,Zheng2021}.

The structure of monolayer MoSSe is shown in Figs.~\ref{MoSSe}(a, b), which is similar to H-MoS$_2$ but with a whole layer of S replaced by Se.
The crystal has space group $P3m1$ and polar point group $C_{3v}$. It follows that
the intrinsic NPHE response obeys Eq.~(\ref{Cnv}), with a single independent tensor element $\chi_{yxxx}^{\text{int}}$.
We evaluate $\chi_{yxxx}^{\text{int}}$ by combining our theory with
first-principles calculations (details are presented in \cite{supp}). The calculated band structure in
Fig.~\ref{MoSSe}(d) matches well with previous
results~\cite{lu2017Janus}. The system is a semiconductor with a
direct gap $\sim 1.47$ eV at $K$ and $K'$. The degeneracy of the top two
valence bands is lifted by a sizable SOC with a large splitting $\sim 168$ meV at $K$. More important for our discussion here is a Rashba-type SOC due to the
the broken horizontal mirror, a characteristic of the Janus structure.
This manifests as a Rashba-type splitting at the $\Gamma$ point~\cite{Chang2021}, as indicated by the arrow in Fig.~\ref{MoSSe}(d) and more clearly in the enlarged Fig.~\ref{MoSSe}(e). We find that this band degeneracy point at $\Gamma$ gives a sizable contribution to the intrinsic NPHE.

The calculated $\chi_{yxxx}^{\text{int}}$ as a function of chemical potential $\mu$ for the $p$-doped case is shown in Fig.~\ref{MoSSe}(f).
Indeed, pronounced response is found when $\mu$ is located around the Rashba-type band degeneracy point at $\Gamma$.
From the calculation, under an in-plane $B$ field of 1 $\text{T}$ the intrinsic nonlinear Hall conductivity can reach $\sim3\times10^{-4}$ $(\Omega\text{V})^{-1}$ if converted to the conventional 3D unit (by dividing the layer thickness of 0.32 nm), when the Fermi level is about 80 meV below the valence band maximum. This value is comparable to those calculated in CuMnAs \cite{Wang2021} and Mn$_2$Au \cite{Liu2021}, and is two orders of magnitude larger than the experimental result measured in CuMnAs \cite{Godinho2018}, so it should be readily detectable. In practice, the carrier doping can be conveniently controlled for 2D materials by electric gating~\cite{Xie2010,Ma2019}. Therefore, our predicted NPHE and its special angular dependence (\ref{Cnv}) (see Fig.~\ref{fig-fig1}(b)) can be directly tested in experiment.

\textcolor{blue}{\textit{Discussion.}} We have proposed a new intrinsic transport effect along with a new intrinsic material property, i.e., the intrinsic NPHE conductivity $\chi^\text{int}$.  It offers a promising characterization tool for a large class of materials, especially the polar and chiral crystals, most of which do not support the linear PHE nor the nonlinear anomalous Hall effect from Berry curvature dipole. Actually, the monolayer MoSSe studied here represents such an example.

Our subject here is on the intrinsic NPHE. As mentioned, there are also other extrinsic contributions. In experiment, they can be separated according to their different scaling with $\tau$ (e.g., by plotting against the longitudinal conductivity)~\cite{Kang2019,Lai2021}. In addition, we point out that the extrinsic contribution is in general not purely a Hall response, namely, the response tensor
$\chi^\text{ext}_{abcd}$ is not guaranteed to be antisymmetric in $a$ and $b$. It follows that their angular dependence will generally differ from that of intrinsic NPHE.

\bibliographystyle{apsrev4-2}
\bibliography{NLPHE_ref}

\begin{thebibliography}{49}%
\makeatletter
\providecommand \@ifxundefined [1]{%
 \@ifx{#1\undefined}
}%
\providecommand \@ifnum [1]{%
 \ifnum #1\expandafter \@firstoftwo
 \else \expandafter \@secondoftwo
 \fi
}%
\providecommand \@ifx [1]{%
 \ifx #1\expandafter \@firstoftwo
 \else \expandafter \@secondoftwo
 \fi
}%
\providecommand \natexlab [1]{#1}%
\providecommand \enquote  [1]{``#1''}%
\providecommand \bibnamefont  [1]{#1}%
\providecommand \bibfnamefont [1]{#1}%
\providecommand \citenamefont [1]{#1}%
\providecommand \href@noop [0]{\@secondoftwo}%
\providecommand \href [0]{\begingroup \@sanitize@url \@href}%
\providecommand \@href[1]{\@@startlink{#1}\@@href}%
\providecommand \@@href[1]{\endgroup#1\@@endlink}%
\providecommand \@sanitize@url [0]{\catcode `\\12\catcode `\$12\catcode
  `\&12\catcode `\#12\catcode `\^12\catcode `\_12\catcode `\%12\relax}%
\providecommand \@@startlink[1]{}%
\providecommand \@@endlink[0]{}%
\providecommand \url  [0]{\begingroup\@sanitize@url \@url }%
\providecommand \@url [1]{\endgroup\@href {#1}{\urlprefix }}%
\providecommand \urlprefix  [0]{URL }%
\providecommand \Eprint [0]{\href }%
\providecommand \doibase [0]{https://doi.org/}%
\providecommand \selectlanguage [0]{\@gobble}%
\providecommand \bibinfo  [0]{\@secondoftwo}%
\providecommand \bibfield  [0]{\@secondoftwo}%
\providecommand \translation [1]{[#1]}%
\providecommand \BibitemOpen [0]{}%
\providecommand \bibitemStop [0]{}%
\providecommand \bibitemNoStop [0]{.\EOS\space}%
\providecommand \EOS [0]{\spacefactor3000\relax}%
\providecommand \BibitemShut  [1]{\csname bibitem#1\endcsname}%
\let\auto@bib@innerbib\@empty
\bibitem [{\citenamefont {Jungwirth}\ \emph {et~al.}(2002)\citenamefont
  {Jungwirth}, \citenamefont {Niu},\ and\ \citenamefont
  {MacDonald}}]{Jungwirth2002}%
  \BibitemOpen
  \bibfield  {author} {\bibinfo {author} {\bibfnamefont {T.}~\bibnamefont
  {Jungwirth}}, \bibinfo {author} {\bibfnamefont {Q.}~\bibnamefont {Niu}},\
  and\ \bibinfo {author} {\bibfnamefont {A.~H.}\ \bibnamefont {MacDonald}},\
  }\href {https://doi.org/10.1103/PhysRevLett.88.207208} {\bibfield  {journal}
  {\bibinfo  {journal} {Phys. Rev. Lett.}\ }\textbf {\bibinfo {volume} {88}},\
  \bibinfo {pages} {207208} (\bibinfo {year} {2002})}\BibitemShut {NoStop}%
\bibitem [{\citenamefont {Onoda}\ and\ \citenamefont
  {Nagaosa}(2002)}]{Nagaosa2002}%
  \BibitemOpen
  \bibfield  {author} {\bibinfo {author} {\bibfnamefont {M.}~\bibnamefont
  {Onoda}}\ and\ \bibinfo {author} {\bibfnamefont {N.}~\bibnamefont
  {Nagaosa}},\ }\href {https://doi.org/10.1143/JPSJ.71.19} {\bibfield
  {journal} {\bibinfo  {journal} {J. Phys. Soc. Jpn.}\ }\textbf {\bibinfo
  {volume} {71}},\ \bibinfo {pages} {19} (\bibinfo {year} {2002})}\BibitemShut
  {NoStop}%
\bibitem [{\citenamefont {Nagaosa}\ \emph {et~al.}(2010)\citenamefont
  {Nagaosa}, \citenamefont {Sinova}, \citenamefont {Onoda}, \citenamefont
  {MacDonald},\ and\ \citenamefont {Ong}}]{Nagaosa2010}%
  \BibitemOpen
  \bibfield  {author} {\bibinfo {author} {\bibfnamefont {N.}~\bibnamefont
  {Nagaosa}}, \bibinfo {author} {\bibfnamefont {J.}~\bibnamefont {Sinova}},
  \bibinfo {author} {\bibfnamefont {S.}~\bibnamefont {Onoda}}, \bibinfo
  {author} {\bibfnamefont {A.~H.}\ \bibnamefont {MacDonald}},\ and\ \bibinfo
  {author} {\bibfnamefont {N.~P.}\ \bibnamefont {Ong}},\ }\href
  {https://doi.org/10.1103/RevModPhys.82.1539} {\bibfield  {journal} {\bibinfo
  {journal} {Rev. Mod. Phys.}\ }\textbf {\bibinfo {volume} {82}},\ \bibinfo
  {pages} {1539} (\bibinfo {year} {2010})}\BibitemShut {NoStop}%
\bibitem [{\citenamefont {Murakami}\ \emph {et~al.}(2003)\citenamefont
  {Murakami}, \citenamefont {Nagaosa},\ and\ \citenamefont
  {Zhang}}]{murakami_dissipationless_2003}%
  \BibitemOpen
  \bibfield  {author} {\bibinfo {author} {\bibfnamefont {S.}~\bibnamefont
  {Murakami}}, \bibinfo {author} {\bibfnamefont {N.}~\bibnamefont {Nagaosa}},\
  and\ \bibinfo {author} {\bibfnamefont {S.-C.}\ \bibnamefont {Zhang}},\ }\href
  {https://doi.org/10.1126/science.1087128} {\bibfield  {journal} {\bibinfo
  {journal} {Science}\ }\textbf {\bibinfo {volume} {301}},\ \bibinfo {pages}
  {1348} (\bibinfo {year} {2003})}\BibitemShut {NoStop}%
\bibitem [{\citenamefont {Sinova}\ \emph {et~al.}(2004)\citenamefont {Sinova},
  \citenamefont {Culcer}, \citenamefont {Niu}, \citenamefont {Sinitsyn},
  \citenamefont {Jungwirth},\ and\ \citenamefont
  {MacDonald}}]{sinova_universal_2004}%
  \BibitemOpen
  \bibfield  {author} {\bibinfo {author} {\bibfnamefont {J.}~\bibnamefont
  {Sinova}}, \bibinfo {author} {\bibfnamefont {D.}~\bibnamefont {Culcer}},
  \bibinfo {author} {\bibfnamefont {Q.}~\bibnamefont {Niu}}, \bibinfo {author}
  {\bibfnamefont {N.~A.}\ \bibnamefont {Sinitsyn}}, \bibinfo {author}
  {\bibfnamefont {T.}~\bibnamefont {Jungwirth}},\ and\ \bibinfo {author}
  {\bibfnamefont {A.~H.}\ \bibnamefont {MacDonald}},\ }\href
  {https://doi.org/10.1103/PhysRevLett.92.126603} {\bibfield  {journal}
  {\bibinfo  {journal} {Phys. Rev. Lett.}\ }\textbf {\bibinfo {volume} {92}},\
  \bibinfo {pages} {126603} (\bibinfo {year} {2004})}\BibitemShut {NoStop}%
\bibitem [{\citenamefont {Sinova}\ \emph {et~al.}(2015)\citenamefont {Sinova},
  \citenamefont {Valenzuela}, \citenamefont {Wunderlich}, \citenamefont
  {Back},\ and\ \citenamefont {Jungwirth}}]{Sinova2015}%
  \BibitemOpen
  \bibfield  {author} {\bibinfo {author} {\bibfnamefont {J.}~\bibnamefont
  {Sinova}}, \bibinfo {author} {\bibfnamefont {S.~O.}\ \bibnamefont
  {Valenzuela}}, \bibinfo {author} {\bibfnamefont {J.}~\bibnamefont
  {Wunderlich}}, \bibinfo {author} {\bibfnamefont {C.~H.}\ \bibnamefont
  {Back}},\ and\ \bibinfo {author} {\bibfnamefont {T.}~\bibnamefont
  {Jungwirth}},\ }\href {https://doi.org/10.1103/RevModPhys.87.1213} {\bibfield
   {journal} {\bibinfo  {journal} {Rev. Mod. Phys.}\ }\textbf {\bibinfo
  {volume} {87}},\ \bibinfo {pages} {1213} (\bibinfo {year}
  {2015})}\BibitemShut {NoStop}%
\bibitem [{\citenamefont {Xiao}\ \emph {et~al.}(2006)\citenamefont {Xiao},
  \citenamefont {Yao}, \citenamefont {Fang},\ and\ \citenamefont
  {Niu}}]{xiao_berry-phase_2006}%
  \BibitemOpen
  \bibfield  {author} {\bibinfo {author} {\bibfnamefont {D.}~\bibnamefont
  {Xiao}}, \bibinfo {author} {\bibfnamefont {Y.}~\bibnamefont {Yao}}, \bibinfo
  {author} {\bibfnamefont {Z.}~\bibnamefont {Fang}},\ and\ \bibinfo {author}
  {\bibfnamefont {Q.}~\bibnamefont {Niu}},\ }\href
  {https://doi.org/10.1103/PhysRevLett.97.026603} {\bibfield  {journal}
  {\bibinfo  {journal} {Phys. Rev. Lett.}\ }\textbf {\bibinfo {volume} {97}},\
  \bibinfo {pages} {026603} (\bibinfo {year} {2006})}\BibitemShut {NoStop}%
\bibitem [{\citenamefont {Xiao}\ \emph {et~al.}(2010)\citenamefont {Xiao},
  \citenamefont {Chang},\ and\ \citenamefont {Niu}}]{Xiao2010}%
  \BibitemOpen
  \bibfield  {author} {\bibinfo {author} {\bibfnamefont {D.}~\bibnamefont
  {Xiao}}, \bibinfo {author} {\bibfnamefont {M.-C.}\ \bibnamefont {Chang}},\
  and\ \bibinfo {author} {\bibfnamefont {Q.}~\bibnamefont {Niu}},\ }\href
  {https://doi.org/10.1103/RevModPhys.82.1959} {\bibfield  {journal} {\bibinfo
  {journal} {Rev. Mod. Phys.}\ }\textbf {\bibinfo {volume} {82}},\ \bibinfo
  {pages} {1959} (\bibinfo {year} {2010})}\BibitemShut {NoStop}%
\bibitem [{\citenamefont {Tang}\ \emph {et~al.}(2003)\citenamefont {Tang},
  \citenamefont {Kawakami}, \citenamefont {Awschalom},\ and\ \citenamefont
  {Roukes}}]{tang_giant_2003}%
  \BibitemOpen
  \bibfield  {author} {\bibinfo {author} {\bibfnamefont {H.~X.}\ \bibnamefont
  {Tang}}, \bibinfo {author} {\bibfnamefont {R.~K.}\ \bibnamefont {Kawakami}},
  \bibinfo {author} {\bibfnamefont {D.~D.}\ \bibnamefont {Awschalom}},\ and\
  \bibinfo {author} {\bibfnamefont {M.~L.}\ \bibnamefont {Roukes}},\ }\href
  {https://doi.org/10.1103/PhysRevLett.90.107201} {\bibfield  {journal}
  {\bibinfo  {journal} {Phys. Rev. Lett.}\ }\textbf {\bibinfo {volume} {90}},\
  \bibinfo {pages} {107201} (\bibinfo {year} {2003})}\BibitemShut {NoStop}%
\bibitem [{\citenamefont {Bowen}\ \emph {et~al.}(2005)\citenamefont {Bowen},
  \citenamefont {Friedland}, \citenamefont {Herfort}, \citenamefont
  {Sch\"{o}nherr},\ and\ \citenamefont {Ploog}}]{bowen_order-driven_2005}%
  \BibitemOpen
  \bibfield  {author} {\bibinfo {author} {\bibfnamefont {M.}~\bibnamefont
  {Bowen}}, \bibinfo {author} {\bibfnamefont {K.-J.}\ \bibnamefont
  {Friedland}}, \bibinfo {author} {\bibfnamefont {J.}~\bibnamefont {Herfort}},
  \bibinfo {author} {\bibfnamefont {H.-P.}\ \bibnamefont {Sch\"{o}nherr}},\
  and\ \bibinfo {author} {\bibfnamefont {K.~H.}\ \bibnamefont {Ploog}},\ }\href
  {https://doi.org/10.1103/PhysRevB.71.172401} {\bibfield  {journal} {\bibinfo
  {journal} {Phys. Rev. B}\ }\textbf {\bibinfo {volume} {71}},\ \bibinfo
  {pages} {172401} (\bibinfo {year} {2005})}\BibitemShut {NoStop}%
\bibitem [{\citenamefont {Seemann}\ \emph {et~al.}(2011)\citenamefont
  {Seemann}, \citenamefont {Freimuth}, \citenamefont {Zhang}, \citenamefont
  {Bl\"{u}gel}, \citenamefont {Mokrousov}, \citenamefont {B\"{u}rgler},\ and\
  \citenamefont {Schneider}}]{seemann_origin_2011}%
  \BibitemOpen
  \bibfield  {author} {\bibinfo {author} {\bibfnamefont {K.~M.}\ \bibnamefont
  {Seemann}}, \bibinfo {author} {\bibfnamefont {F.}~\bibnamefont {Freimuth}},
  \bibinfo {author} {\bibfnamefont {H.}~\bibnamefont {Zhang}}, \bibinfo
  {author} {\bibfnamefont {S.}~\bibnamefont {Bl\"{u}gel}}, \bibinfo {author}
  {\bibfnamefont {Y.}~\bibnamefont {Mokrousov}}, \bibinfo {author}
  {\bibfnamefont {D.~E.}\ \bibnamefont {B\"{u}rgler}},\ and\ \bibinfo {author}
  {\bibfnamefont {C.~M.}\ \bibnamefont {Schneider}},\ }\href
  {https://doi.org/10.1103/PhysRevLett.107.086603} {\bibfield  {journal}
  {\bibinfo  {journal} {Phys. Rev. Lett.}\ }\textbf {\bibinfo {volume} {107}},\
  \bibinfo {pages} {086603} (\bibinfo {year} {2011})}\BibitemShut {NoStop}%
\bibitem [{\citenamefont {Nandy}\ \emph {et~al.}(2017)\citenamefont {Nandy},
  \citenamefont {Sharma}, \citenamefont {Taraphder},\ and\ \citenamefont
  {Tewari}}]{Nandy2017}%
  \BibitemOpen
  \bibfield  {author} {\bibinfo {author} {\bibfnamefont {S.}~\bibnamefont
  {Nandy}}, \bibinfo {author} {\bibfnamefont {G.}~\bibnamefont {Sharma}},
  \bibinfo {author} {\bibfnamefont {A.}~\bibnamefont {Taraphder}},\ and\
  \bibinfo {author} {\bibfnamefont {S.}~\bibnamefont {Tewari}},\ }\href
  {https://doi.org/10.1103/PhysRevLett.119.176804} {\bibfield  {journal}
  {\bibinfo  {journal} {Phys. Rev. Lett.}\ }\textbf {\bibinfo {volume} {119}},\
  \bibinfo {pages} {176804} (\bibinfo {year} {2017})}\BibitemShut {NoStop}%
\bibitem [{\citenamefont {Kumar}\ \emph {et~al.}(2018)\citenamefont {Kumar},
  \citenamefont {Guin}, \citenamefont {Felser},\ and\ \citenamefont
  {Shekhar}}]{kumar_planar_2018}%
  \BibitemOpen
  \bibfield  {author} {\bibinfo {author} {\bibfnamefont {N.}~\bibnamefont
  {Kumar}}, \bibinfo {author} {\bibfnamefont {S.~N.}\ \bibnamefont {Guin}},
  \bibinfo {author} {\bibfnamefont {C.}~\bibnamefont {Felser}},\ and\ \bibinfo
  {author} {\bibfnamefont {C.}~\bibnamefont {Shekhar}},\ }\href
  {https://doi.org/10.1103/PhysRevB.98.041103} {\bibfield  {journal} {\bibinfo
  {journal} {Phys. Rev. B}\ }\textbf {\bibinfo {volume} {98}},\ \bibinfo
  {pages} {041103(R)} (\bibinfo {year} {2018})}\BibitemShut {NoStop}%
\bibitem [{\citenamefont {Deng}\ \emph {et~al.}(2019)\citenamefont {Deng},
  \citenamefont {Duan}, \citenamefont {Luo}, \citenamefont {Deng},
  \citenamefont {Wang},\ and\ \citenamefont {Sheng}}]{deng_quantum_2019}%
  \BibitemOpen
  \bibfield  {author} {\bibinfo {author} {\bibfnamefont {M.-X.}\ \bibnamefont
  {Deng}}, \bibinfo {author} {\bibfnamefont {H.-J.}\ \bibnamefont {Duan}},
  \bibinfo {author} {\bibfnamefont {W.}~\bibnamefont {Luo}}, \bibinfo {author}
  {\bibfnamefont {W.~Y.}\ \bibnamefont {Deng}}, \bibinfo {author}
  {\bibfnamefont {R.-Q.}\ \bibnamefont {Wang}},\ and\ \bibinfo {author}
  {\bibfnamefont {L.}~\bibnamefont {Sheng}},\ }\href
  {https://doi.org/10.1103/PhysRevB.99.165146} {\bibfield  {journal} {\bibinfo
  {journal} {Phys. Rev. B}\ }\textbf {\bibinfo {volume} {99}},\ \bibinfo
  {pages} {165146} (\bibinfo {year} {2019})}\BibitemShut {NoStop}%
\bibitem [{\citenamefont {Ma}\ \emph {et~al.}(2019{\natexlab{a}})\citenamefont
  {Ma}, \citenamefont {Jiang}, \citenamefont {Liu},\ and\ \citenamefont
  {Xie}}]{ma_planar_2019}%
  \BibitemOpen
  \bibfield  {author} {\bibinfo {author} {\bibfnamefont {D.}~\bibnamefont
  {Ma}}, \bibinfo {author} {\bibfnamefont {H.}~\bibnamefont {Jiang}}, \bibinfo
  {author} {\bibfnamefont {H.}~\bibnamefont {Liu}},\ and\ \bibinfo {author}
  {\bibfnamefont {X.~C.}\ \bibnamefont {Xie}},\ }\href
  {https://doi.org/10.1103/PhysRevB.99.115121} {\bibfield  {journal} {\bibinfo
  {journal} {Phys. Rev. B}\ }\textbf {\bibinfo {volume} {99}},\ \bibinfo
  {pages} {115121} (\bibinfo {year} {2019}{\natexlab{a}})}\BibitemShut
  {NoStop}%
\bibitem [{\citenamefont {Breunig}\ \emph {et~al.}(2017)\citenamefont
  {Breunig}, \citenamefont {Wang}, \citenamefont {Taskin}, \citenamefont {Lux},
  \citenamefont {Rosch},\ and\ \citenamefont {Ando}}]{breunig_gigantic_2017}%
  \BibitemOpen
  \bibfield  {author} {\bibinfo {author} {\bibfnamefont {O.}~\bibnamefont
  {Breunig}}, \bibinfo {author} {\bibfnamefont {Z.}~\bibnamefont {Wang}},
  \bibinfo {author} {\bibfnamefont {A.~A.}\ \bibnamefont {Taskin}}, \bibinfo
  {author} {\bibfnamefont {J.}~\bibnamefont {Lux}}, \bibinfo {author}
  {\bibfnamefont {A.}~\bibnamefont {Rosch}},\ and\ \bibinfo {author}
  {\bibfnamefont {Y.}~\bibnamefont {Ando}},\ }\href
  {https://doi.org/10.1038/ncomms15545} {\bibfield  {journal} {\bibinfo
  {journal} {Nat. Commun.}\ }\textbf {\bibinfo {volume} {8}},\ \bibinfo {pages}
  {15545} (\bibinfo {year} {2017})}\BibitemShut {NoStop}%
\bibitem [{\citenamefont {Wu}\ \emph {et~al.}(2018{\natexlab{a}})\citenamefont
  {Wu}, \citenamefont {Pan}, \citenamefont {Wu}, \citenamefont {Fei},
  \citenamefont {Chen}, \citenamefont {Liu}, \citenamefont {Bu}, \citenamefont
  {Cao}, \citenamefont {Song},\ and\ \citenamefont
  {Wang}}]{wu_oscillating_2018}%
  \BibitemOpen
  \bibfield  {author} {\bibinfo {author} {\bibfnamefont {B.}~\bibnamefont
  {Wu}}, \bibinfo {author} {\bibfnamefont {X.-C.}\ \bibnamefont {Pan}},
  \bibinfo {author} {\bibfnamefont {W.}~\bibnamefont {Wu}}, \bibinfo {author}
  {\bibfnamefont {F.}~\bibnamefont {Fei}}, \bibinfo {author} {\bibfnamefont
  {B.}~\bibnamefont {Chen}}, \bibinfo {author} {\bibfnamefont {Q.}~\bibnamefont
  {Liu}}, \bibinfo {author} {\bibfnamefont {H.}~\bibnamefont {Bu}}, \bibinfo
  {author} {\bibfnamefont {L.}~\bibnamefont {Cao}}, \bibinfo {author}
  {\bibfnamefont {F.}~\bibnamefont {Song}},\ and\ \bibinfo {author}
  {\bibfnamefont {B.}~\bibnamefont {Wang}},\ }\href
  {https://doi.org/10.1063/1.5031906} {\bibfield  {journal} {\bibinfo
  {journal} {Appl. Phys. Lett.}\ }\textbf {\bibinfo {volume} {113}},\ \bibinfo
  {pages} {011902} (\bibinfo {year} {2018}{\natexlab{a}})}\BibitemShut
  {NoStop}%
\bibitem [{\citenamefont {Rakhmilevich}\ \emph {et~al.}(2018)\citenamefont
  {Rakhmilevich}, \citenamefont {Wang}, \citenamefont {Zhao}, \citenamefont
  {Chan}, \citenamefont {Moodera}, \citenamefont {Liu},\ and\ \citenamefont
  {Chang}}]{rakhmilevich_unconventional_2018}%
  \BibitemOpen
  \bibfield  {author} {\bibinfo {author} {\bibfnamefont {D.}~\bibnamefont
  {Rakhmilevich}}, \bibinfo {author} {\bibfnamefont {F.}~\bibnamefont {Wang}},
  \bibinfo {author} {\bibfnamefont {W.}~\bibnamefont {Zhao}}, \bibinfo {author}
  {\bibfnamefont {M.~H.~W.}\ \bibnamefont {Chan}}, \bibinfo {author}
  {\bibfnamefont {J.~S.}\ \bibnamefont {Moodera}}, \bibinfo {author}
  {\bibfnamefont {C.}~\bibnamefont {Liu}},\ and\ \bibinfo {author}
  {\bibfnamefont {C.-Z.}\ \bibnamefont {Chang}},\ }\href
  {https://doi.org/10.1103/PhysRevB.98.094404} {\bibfield  {journal} {\bibinfo
  {journal} {Phys. Rev. B}\ }\textbf {\bibinfo {volume} {98}},\ \bibinfo
  {pages} {094404} (\bibinfo {year} {2018})}\BibitemShut {NoStop}%
\bibitem [{\citenamefont {He}\ \emph {et~al.}(2019)\citenamefont {He},
  \citenamefont {Zhang}, \citenamefont {Zhu}, \citenamefont {Shi},
  \citenamefont {Heinonen}, \citenamefont {Vignale},\ and\ \citenamefont
  {Yang}}]{He2019}%
  \BibitemOpen
  \bibfield  {author} {\bibinfo {author} {\bibfnamefont {P.}~\bibnamefont
  {He}}, \bibinfo {author} {\bibfnamefont {S.~S.-L.}\ \bibnamefont {Zhang}},
  \bibinfo {author} {\bibfnamefont {D.}~\bibnamefont {Zhu}}, \bibinfo {author}
  {\bibfnamefont {S.}~\bibnamefont {Shi}}, \bibinfo {author} {\bibfnamefont
  {O.~G.}\ \bibnamefont {Heinonen}}, \bibinfo {author} {\bibfnamefont
  {G.}~\bibnamefont {Vignale}},\ and\ \bibinfo {author} {\bibfnamefont
  {H.}~\bibnamefont {Yang}},\ }\href
  {https://doi.org/10.1103/PhysRevLett.123.016801} {\bibfield  {journal}
  {\bibinfo  {journal} {Phys. Rev. Lett.}\ }\textbf {\bibinfo {volume} {123}},\
  \bibinfo {pages} {016801} (\bibinfo {year} {2019})}\BibitemShut {NoStop}%
\bibitem [{\citenamefont {Esin}\ \emph {et~al.}(2021)\citenamefont {Esin},
  \citenamefont {Timonina}, \citenamefont {Kolesnikov},\ and\ \citenamefont
  {Deviatov}}]{esin_nonlinear_2021}%
  \BibitemOpen
  \bibfield  {author} {\bibinfo {author} {\bibfnamefont {V.~D.}\ \bibnamefont
  {Esin}}, \bibinfo {author} {\bibfnamefont {A.~V.}\ \bibnamefont {Timonina}},
  \bibinfo {author} {\bibfnamefont {N.~N.}\ \bibnamefont {Kolesnikov}},\ and\
  \bibinfo {author} {\bibfnamefont {E.~V.}\ \bibnamefont {Deviatov}},\ }\href
  {https://doi.org/10.1134/S1063776121120037} {\bibfield  {journal} {\bibinfo
  {journal} {J. Exp. Theor. Phys.}\ }\textbf {\bibinfo {volume} {133}},\
  \bibinfo {pages} {792} (\bibinfo {year} {2021})}\BibitemShut {NoStop}%
\bibitem [{\citenamefont {Zheng}\ \emph {et~al.}(2020)\citenamefont {Zheng},
  \citenamefont {Duan}, \citenamefont {Wang}, \citenamefont {Li}, \citenamefont
  {Deng},\ and\ \citenamefont {Wang}}]{zheng_origin_2020}%
  \BibitemOpen
  \bibfield  {author} {\bibinfo {author} {\bibfnamefont {S.-H.}\ \bibnamefont
  {Zheng}}, \bibinfo {author} {\bibfnamefont {H.-J.}\ \bibnamefont {Duan}},
  \bibinfo {author} {\bibfnamefont {J.-K.}\ \bibnamefont {Wang}}, \bibinfo
  {author} {\bibfnamefont {J.-Y.}\ \bibnamefont {Li}}, \bibinfo {author}
  {\bibfnamefont {M.-X.}\ \bibnamefont {Deng}},\ and\ \bibinfo {author}
  {\bibfnamefont {R.-Q.}\ \bibnamefont {Wang}},\ }\href
  {https://doi.org/10.1103/PhysRevB.101.041408} {\bibfield  {journal} {\bibinfo
   {journal} {Phys. Rev. B}\ }\textbf {\bibinfo {volume} {101}},\ \bibinfo
  {pages} {041408(R)} (\bibinfo {year} {2020})}\BibitemShut {NoStop}%
\bibitem [{\citenamefont {Rao}\ \emph {et~al.}(2021)\citenamefont {Rao},
  \citenamefont {Zhou}, \citenamefont {Wu}, \citenamefont {Duan}, \citenamefont
  {Deng},\ and\ \citenamefont {Wang}}]{rao_theory_2021}%
  \BibitemOpen
  \bibfield  {author} {\bibinfo {author} {\bibfnamefont {W.}~\bibnamefont
  {Rao}}, \bibinfo {author} {\bibfnamefont {Y.-L.}\ \bibnamefont {Zhou}},
  \bibinfo {author} {\bibfnamefont {Y.-j.}\ \bibnamefont {Wu}}, \bibinfo
  {author} {\bibfnamefont {H.-J.}\ \bibnamefont {Duan}}, \bibinfo {author}
  {\bibfnamefont {M.-X.}\ \bibnamefont {Deng}},\ and\ \bibinfo {author}
  {\bibfnamefont {R.-Q.}\ \bibnamefont {Wang}},\ }\href
  {https://doi.org/10.1103/PhysRevB.103.155415} {\bibfield  {journal} {\bibinfo
   {journal} {Phys. Rev. B}\ }\textbf {\bibinfo {volume} {103}},\ \bibinfo
  {pages} {155415} (\bibinfo {year} {2021})}\BibitemShut {NoStop}%
\bibitem [{\citenamefont {Battilomo}\ \emph {et~al.}(2021)\citenamefont
  {Battilomo}, \citenamefont {Scopigno},\ and\ \citenamefont
  {Ortix}}]{battilomo_anomalous_2021}%
  \BibitemOpen
  \bibfield  {author} {\bibinfo {author} {\bibfnamefont {R.}~\bibnamefont
  {Battilomo}}, \bibinfo {author} {\bibfnamefont {N.}~\bibnamefont
  {Scopigno}},\ and\ \bibinfo {author} {\bibfnamefont {C.}~\bibnamefont
  {Ortix}},\ }\href {https://doi.org/10.1103/PhysRevResearch.3.L012006}
  {\bibfield  {journal} {\bibinfo  {journal} {Phys. Rev. Res.}\ }\textbf
  {\bibinfo {volume} {3}},\ \bibinfo {pages} {L012006} (\bibinfo {year}
  {2021})}\BibitemShut {NoStop}%
\bibitem [{\citenamefont {Gao}\ \emph {et~al.}(2014)\citenamefont {Gao},
  \citenamefont {Yang},\ and\ \citenamefont {Niu}}]{Gao2014}%
  \BibitemOpen
  \bibfield  {author} {\bibinfo {author} {\bibfnamefont {Y.}~\bibnamefont
  {Gao}}, \bibinfo {author} {\bibfnamefont {S.~A.}\ \bibnamefont {Yang}},\ and\
  \bibinfo {author} {\bibfnamefont {Q.}~\bibnamefont {Niu}},\ }\href
  {https://doi.org/10.1103/PhysRevLett.112.166601} {\bibfield  {journal}
  {\bibinfo  {journal} {Phys. Rev. Lett.}\ }\textbf {\bibinfo {volume} {112}},\
  \bibinfo {pages} {166601} (\bibinfo {year} {2014})}\BibitemShut {NoStop}%
\bibitem [{\citenamefont {Liu}\ \emph {et~al.}(2022)\citenamefont {Liu},
  \citenamefont {Zhao}, \citenamefont {Huang}, \citenamefont {Feng},
  \citenamefont {Xiao}, \citenamefont {Wu}, \citenamefont {Lai}, \citenamefont
  {Gao},\ and\ \citenamefont {Yang}}]{Liu2022}%
  \BibitemOpen
  \bibfield  {author} {\bibinfo {author} {\bibfnamefont {H.}~\bibnamefont
  {Liu}}, \bibinfo {author} {\bibfnamefont {J.}~\bibnamefont {Zhao}}, \bibinfo
  {author} {\bibfnamefont {Y.-X.}\ \bibnamefont {Huang}}, \bibinfo {author}
  {\bibfnamefont {X.}~\bibnamefont {Feng}}, \bibinfo {author} {\bibfnamefont
  {C.}~\bibnamefont {Xiao}}, \bibinfo {author} {\bibfnamefont {W.}~\bibnamefont
  {Wu}}, \bibinfo {author} {\bibfnamefont {S.}~\bibnamefont {Lai}}, \bibinfo
  {author} {\bibfnamefont {W.-b.}\ \bibnamefont {Gao}},\ and\ \bibinfo {author}
  {\bibfnamefont {S.~A.}\ \bibnamefont {Yang}},\ }\href
  {https://doi.org/10.1103/PhysRevB.105.045118} {\bibfield  {journal} {\bibinfo
   {journal} {Phys. Rev. B}\ }\textbf {\bibinfo {volume} {105}},\ \bibinfo
  {pages} {045118} (\bibinfo {year} {2022})}\BibitemShut {NoStop}%
\bibitem [{\citenamefont {Sundaram}\ and\ \citenamefont
  {Niu}(1999)}]{sundaram_wave-packet_1999}%
  \BibitemOpen
  \bibfield  {author} {\bibinfo {author} {\bibfnamefont {G.}~\bibnamefont
  {Sundaram}}\ and\ \bibinfo {author} {\bibfnamefont {Q.}~\bibnamefont {Niu}},\
  }\href {https://doi.org/10.1103/PhysRevB.59.14915} {\bibfield  {journal}
  {\bibinfo  {journal} {Phys. Rev. B}\ }\textbf {\bibinfo {volume} {59}},\
  \bibinfo {pages} {14915} (\bibinfo {year} {1999})}\BibitemShut {NoStop}%
\bibitem [{\citenamefont {Karplus}\ and\ \citenamefont
  {Luttinger}(1954)}]{karplus_hall_1954}%
  \BibitemOpen
  \bibfield  {author} {\bibinfo {author} {\bibfnamefont {R.}~\bibnamefont
  {Karplus}}\ and\ \bibinfo {author} {\bibfnamefont {J.~M.}\ \bibnamefont
  {Luttinger}},\ }\href {https://doi.org/10.1103/PhysRev.95.1154} {\bibfield
  {journal} {\bibinfo  {journal} {Phys. Rev.}\ }\textbf {\bibinfo {volume}
  {95}},\ \bibinfo {pages} {1154} (\bibinfo {year} {1954})}\BibitemShut
  {NoStop}%
\bibitem [{\citenamefont {Chang}\ and\ \citenamefont
  {Niu}(1995)}]{chang_berry_1995}%
  \BibitemOpen
  \bibfield  {author} {\bibinfo {author} {\bibfnamefont {M.-C.}\ \bibnamefont
  {Chang}}\ and\ \bibinfo {author} {\bibfnamefont {Q.}~\bibnamefont {Niu}},\
  }\href {https://doi.org/10.1103/PhysRevLett.75.1348} {\bibfield  {journal}
  {\bibinfo  {journal} {Phys. Rev. Lett.}\ }\textbf {\bibinfo {volume} {75}},\
  \bibinfo {pages} {1348} (\bibinfo {year} {1995})}\BibitemShut {NoStop}%
\bibitem [{\citenamefont {Gao}\ \emph {et~al.}(2015)\citenamefont {Gao},
  \citenamefont {Yang},\ and\ \citenamefont {Niu}}]{Gao2015}%
  \BibitemOpen
  \bibfield  {author} {\bibinfo {author} {\bibfnamefont {Y.}~\bibnamefont
  {Gao}}, \bibinfo {author} {\bibfnamefont {S.~A.}\ \bibnamefont {Yang}},\ and\
  \bibinfo {author} {\bibfnamefont {Q.}~\bibnamefont {Niu}},\ }\href
  {https://doi.org/10.1103/PhysRevB.91.214405} {\bibfield  {journal} {\bibinfo
  {journal} {Phys. Rev. B}\ }\textbf {\bibinfo {volume} {91}},\ \bibinfo
  {pages} {214405} (\bibinfo {year} {2015})}\BibitemShut {NoStop}%
\bibitem [{sup()}]{supp}%
  \BibitemOpen
  \href@noop {} {}\bibinfo {howpublished} {See Supplemental Material for
  theoretical and calculation details.}\BibitemShut {Stop}%
\bibitem [{\citenamefont {Kang}\ \emph {et~al.}(2019)\citenamefont {Kang},
  \citenamefont {Li}, \citenamefont {Sohn}, \citenamefont {Shan},\ and\
  \citenamefont {Mak}}]{Kang2019}%
  \BibitemOpen
  \bibfield  {author} {\bibinfo {author} {\bibfnamefont {K.}~\bibnamefont
  {Kang}}, \bibinfo {author} {\bibfnamefont {T.}~\bibnamefont {Li}}, \bibinfo
  {author} {\bibfnamefont {E.}~\bibnamefont {Sohn}}, \bibinfo {author}
  {\bibfnamefont {J.}~\bibnamefont {Shan}},\ and\ \bibinfo {author}
  {\bibfnamefont {K.~F.}\ \bibnamefont {Mak}},\ }\href
  {https://doi.org/10.1038/s41563-019-0294-7} {\bibfield  {journal} {\bibinfo
  {journal} {Nat. Mater.}\ }\textbf {\bibinfo {volume} {18}},\ \bibinfo {pages}
  {324} (\bibinfo {year} {2019})}\BibitemShut {NoStop}%
\bibitem [{\citenamefont {Lai}\ \emph {et~al.}(2021)\citenamefont {Lai},
  \citenamefont {Liu}, \citenamefont {Zhang}, \citenamefont {Zhao},
  \citenamefont {Feng}, \citenamefont {Wang}, \citenamefont {Tang},
  \citenamefont {Liu}, \citenamefont {Novoselov}, \citenamefont {Yang},\ and\
  \citenamefont {Gao}}]{Lai2021}%
  \BibitemOpen
  \bibfield  {author} {\bibinfo {author} {\bibfnamefont {S.}~\bibnamefont
  {Lai}}, \bibinfo {author} {\bibfnamefont {H.}~\bibnamefont {Liu}}, \bibinfo
  {author} {\bibfnamefont {Z.}~\bibnamefont {Zhang}}, \bibinfo {author}
  {\bibfnamefont {J.}~\bibnamefont {Zhao}}, \bibinfo {author} {\bibfnamefont
  {X.}~\bibnamefont {Feng}}, \bibinfo {author} {\bibfnamefont {N.}~\bibnamefont
  {Wang}}, \bibinfo {author} {\bibfnamefont {C.}~\bibnamefont {Tang}}, \bibinfo
  {author} {\bibfnamefont {Y.}~\bibnamefont {Liu}}, \bibinfo {author}
  {\bibfnamefont {K.~S.}\ \bibnamefont {Novoselov}}, \bibinfo {author}
  {\bibfnamefont {S.~A.}\ \bibnamefont {Yang}},\ and\ \bibinfo {author}
  {\bibfnamefont {W.-b.}\ \bibnamefont {Gao}},\ }\href
  {https://doi.org/10.1038/s41565-021-00917-0} {\bibfield  {journal} {\bibinfo
  {journal} {Nat. Nanotechnol.}\ }\textbf {\bibinfo {volume} {16}},\ \bibinfo
  {pages} {869} (\bibinfo {year} {2021})}\BibitemShut {NoStop}%
\bibitem [{not()}]{note}%
  \BibitemOpen
  \href@noop {} {}\bibinfo {howpublished} {Here, for $C_{2v}$, the twofold axis
  is normal to the plane.}\BibitemShut {Stop}%
\bibitem [{\citenamefont {Sodemann}\ and\ \citenamefont {Fu}(2015)}]{Fu2015}%
  \BibitemOpen
  \bibfield  {author} {\bibinfo {author} {\bibfnamefont {I.}~\bibnamefont
  {Sodemann}}\ and\ \bibinfo {author} {\bibfnamefont {L.}~\bibnamefont {Fu}},\
  }\href {https://doi.org/10.1103/PhysRevLett.115.216806} {\bibfield  {journal}
  {\bibinfo  {journal} {Phys. Rev. Lett.}\ }\textbf {\bibinfo {volume} {115}},\
  \bibinfo {pages} {216806} (\bibinfo {year} {2015})}\BibitemShut {NoStop}%
\bibitem [{\citenamefont {Liu}\ \emph {et~al.}(2011{\natexlab{a}})\citenamefont
  {Liu}, \citenamefont {Jiang},\ and\ \citenamefont
  {Yao}}]{liu_low-energy_2011}%
  \BibitemOpen
  \bibfield  {author} {\bibinfo {author} {\bibfnamefont {C.-C.}\ \bibnamefont
  {Liu}}, \bibinfo {author} {\bibfnamefont {H.}~\bibnamefont {Jiang}},\ and\
  \bibinfo {author} {\bibfnamefont {Y.}~\bibnamefont {Yao}},\ }\href
  {https://doi.org/10.1103/PhysRevB.84.195430} {\bibfield  {journal} {\bibinfo
  {journal} {Phys. Rev. B}\ }\textbf {\bibinfo {volume} {84}},\ \bibinfo
  {pages} {195430} (\bibinfo {year} {2011}{\natexlab{a}})}\BibitemShut
  {NoStop}%
\bibitem [{\citenamefont {Liu}\ \emph {et~al.}(2011{\natexlab{b}})\citenamefont
  {Liu}, \citenamefont {Feng},\ and\ \citenamefont {Yao}}]{liu_quantum_2011}%
  \BibitemOpen
  \bibfield  {author} {\bibinfo {author} {\bibfnamefont {C.-C.}\ \bibnamefont
  {Liu}}, \bibinfo {author} {\bibfnamefont {W.}~\bibnamefont {Feng}},\ and\
  \bibinfo {author} {\bibfnamefont {Y.}~\bibnamefont {Yao}},\ }\href
  {https://doi.org/10.1103/PhysRevLett.107.076802} {\bibfield  {journal}
  {\bibinfo  {journal} {Phys. Rev. Lett.}\ }\textbf {\bibinfo {volume} {107}},\
  \bibinfo {pages} {076802} (\bibinfo {year} {2011}{\natexlab{b}})}\BibitemShut
  {NoStop}%
\bibitem [{\citenamefont {Vanderbilt}(2018)}]{vanderbilt_berry_2018}%
  \BibitemOpen
  \bibfield  {author} {\bibinfo {author} {\bibfnamefont {D.}~\bibnamefont
  {Vanderbilt}},\ }\href@noop {} {\emph {\bibinfo {title} {Berry phases in
  electronic structure theory}}}\ (\bibinfo  {publisher} {Cambridge University
  Press},\ \bibinfo {address} {Cambridge},\ \bibinfo {year} {2018})\BibitemShut
  {NoStop}%
\bibitem [{\citenamefont {Wu}\ \emph {et~al.}(2018{\natexlab{b}})\citenamefont
  {Wu}, \citenamefont {Liu}, \citenamefont {Li}, \citenamefont {Zhong},
  \citenamefont {Yu}, \citenamefont {Sheng}, \citenamefont {Zhao},\ and\
  \citenamefont {Yang}}]{wu_nodal_2018}%
  \BibitemOpen
  \bibfield  {author} {\bibinfo {author} {\bibfnamefont {W.}~\bibnamefont
  {Wu}}, \bibinfo {author} {\bibfnamefont {Y.}~\bibnamefont {Liu}}, \bibinfo
  {author} {\bibfnamefont {S.}~\bibnamefont {Li}}, \bibinfo {author}
  {\bibfnamefont {C.}~\bibnamefont {Zhong}}, \bibinfo {author} {\bibfnamefont
  {Z.-M.}\ \bibnamefont {Yu}}, \bibinfo {author} {\bibfnamefont {X.-L.}\
  \bibnamefont {Sheng}}, \bibinfo {author} {\bibfnamefont {Y.~X.}\ \bibnamefont
  {Zhao}},\ and\ \bibinfo {author} {\bibfnamefont {S.~A.}\ \bibnamefont
  {Yang}},\ }\href {https://doi.org/10.1103/PhysRevB.97.115125} {\bibfield
  {journal} {\bibinfo  {journal} {Phys. Rev. B}\ }\textbf {\bibinfo {volume}
  {97}},\ \bibinfo {pages} {115125} (\bibinfo {year}
  {2018}{\natexlab{b}})}\BibitemShut {NoStop}%
\bibitem [{\citenamefont {Zhang}\ \emph {et~al.}(2017)\citenamefont {Zhang},
  \citenamefont {Jia}, \citenamefont {Kholmanov}, \citenamefont {Dong},
  \citenamefont {Er}, \citenamefont {Chen}, \citenamefont {Guo}, \citenamefont
  {Jin}, \citenamefont {Shenoy}, \citenamefont {Shi} \emph
  {et~al.}}]{zhang2017janus}%
  \BibitemOpen
  \bibfield  {author} {\bibinfo {author} {\bibfnamefont {J.}~\bibnamefont
  {Zhang}}, \bibinfo {author} {\bibfnamefont {S.}~\bibnamefont {Jia}}, \bibinfo
  {author} {\bibfnamefont {I.}~\bibnamefont {Kholmanov}}, \bibinfo {author}
  {\bibfnamefont {L.}~\bibnamefont {Dong}}, \bibinfo {author} {\bibfnamefont
  {D.}~\bibnamefont {Er}}, \bibinfo {author} {\bibfnamefont {W.}~\bibnamefont
  {Chen}}, \bibinfo {author} {\bibfnamefont {H.}~\bibnamefont {Guo}}, \bibinfo
  {author} {\bibfnamefont {Z.}~\bibnamefont {Jin}}, \bibinfo {author}
  {\bibfnamefont {V.~B.}\ \bibnamefont {Shenoy}}, \bibinfo {author}
  {\bibfnamefont {L.}~\bibnamefont {Shi}}, \emph {et~al.},\ }\href
  {https://doi.org/10.1021/acsnano.7b03186} {\bibfield  {journal} {\bibinfo
  {journal} {ACS Nano}\ }\textbf {\bibinfo {volume} {11}},\ \bibinfo {pages}
  {8192} (\bibinfo {year} {2017})}\BibitemShut {NoStop}%
\bibitem [{\citenamefont {Lu}\ \emph {et~al.}(2017)\citenamefont {Lu},
  \citenamefont {Zhu}, \citenamefont {Xiao}, \citenamefont {Chuu},
  \citenamefont {Han}, \citenamefont {Chiu}, \citenamefont {Cheng},
  \citenamefont {Yang}, \citenamefont {Wei}, \citenamefont {Yang} \emph
  {et~al.}}]{lu2017Janus}%
  \BibitemOpen
  \bibfield  {author} {\bibinfo {author} {\bibfnamefont {A.-Y.}\ \bibnamefont
  {Lu}}, \bibinfo {author} {\bibfnamefont {H.}~\bibnamefont {Zhu}}, \bibinfo
  {author} {\bibfnamefont {J.}~\bibnamefont {Xiao}}, \bibinfo {author}
  {\bibfnamefont {C.-P.}\ \bibnamefont {Chuu}}, \bibinfo {author}
  {\bibfnamefont {Y.}~\bibnamefont {Han}}, \bibinfo {author} {\bibfnamefont
  {M.-H.}\ \bibnamefont {Chiu}}, \bibinfo {author} {\bibfnamefont {C.-C.}\
  \bibnamefont {Cheng}}, \bibinfo {author} {\bibfnamefont {C.-W.}\ \bibnamefont
  {Yang}}, \bibinfo {author} {\bibfnamefont {K.-H.}\ \bibnamefont {Wei}},
  \bibinfo {author} {\bibfnamefont {Y.}~\bibnamefont {Yang}}, \emph {et~al.},\
  }\href {https://doi.org/10.1038/nnano.2017.100} {\bibfield  {journal}
  {\bibinfo  {journal} {Nat. Nanotechnol.}\ }\textbf {\bibinfo {volume} {12}},\
  \bibinfo {pages} {744} (\bibinfo {year} {2017})}\BibitemShut {NoStop}%
\bibitem [{\citenamefont {Dong}\ \emph {et~al.}(2017)\citenamefont {Dong},
  \citenamefont {Lou},\ and\ \citenamefont {Shenoy}}]{Dong2017}%
  \BibitemOpen
  \bibfield  {author} {\bibinfo {author} {\bibfnamefont {L.}~\bibnamefont
  {Dong}}, \bibinfo {author} {\bibfnamefont {J.}~\bibnamefont {Lou}},\ and\
  \bibinfo {author} {\bibfnamefont {V.~B.}\ \bibnamefont {Shenoy}},\ }\href
  {https://doi.org/10.1021/acsnano.7b03313} {\bibfield  {journal} {\bibinfo
  {journal} {ACS Nano}\ }\textbf {\bibinfo {volume} {11}},\ \bibinfo {pages}
  {8242} (\bibinfo {year} {2017})}\BibitemShut {NoStop}%
\bibitem [{\citenamefont {Guo}\ \emph {et~al.}(2021)\citenamefont {Guo},
  \citenamefont {Lin}, \citenamefont {Xie}, \citenamefont {Yuan}, \citenamefont
  {Zhu}, \citenamefont {Shen}, \citenamefont {Lu}, \citenamefont {Su},
  \citenamefont {Shi}, \citenamefont {Zhang} \emph {et~al.}}]{Guo2021}%
  \BibitemOpen
  \bibfield  {author} {\bibinfo {author} {\bibfnamefont {Y.}~\bibnamefont
  {Guo}}, \bibinfo {author} {\bibfnamefont {Y.}~\bibnamefont {Lin}}, \bibinfo
  {author} {\bibfnamefont {K.}~\bibnamefont {Xie}}, \bibinfo {author}
  {\bibfnamefont {B.}~\bibnamefont {Yuan}}, \bibinfo {author} {\bibfnamefont
  {J.}~\bibnamefont {Zhu}}, \bibinfo {author} {\bibfnamefont {P.-C.}\
  \bibnamefont {Shen}}, \bibinfo {author} {\bibfnamefont {A.-Y.}\ \bibnamefont
  {Lu}}, \bibinfo {author} {\bibfnamefont {C.}~\bibnamefont {Su}}, \bibinfo
  {author} {\bibfnamefont {E.}~\bibnamefont {Shi}}, \bibinfo {author}
  {\bibfnamefont {K.}~\bibnamefont {Zhang}}, \emph {et~al.},\ }\href
  {https://doi.org/10.1073/pnas.2106124118} {\bibfield  {journal} {\bibinfo
  {journal} {Proc. Natl. Acad. Sci. U.S.A.}\ }\textbf {\bibinfo {volume}
  {118}},\ \bibinfo {pages} {e2106124118} (\bibinfo {year} {2021})}\BibitemShut
  {NoStop}%
\bibitem [{\citenamefont {Zheng}\ \emph {et~al.}(2021)\citenamefont {Zheng},
  \citenamefont {Lin}, \citenamefont {Yu}, \citenamefont {Valencia-Acuna},
  \citenamefont {Puretzky}, \citenamefont {Torsi}, \citenamefont {Liu},
  \citenamefont {Ivanov}, \citenamefont {Duscher}, \citenamefont {Geohegan}
  \emph {et~al.}}]{Zheng2021}%
  \BibitemOpen
  \bibfield  {author} {\bibinfo {author} {\bibfnamefont {T.}~\bibnamefont
  {Zheng}}, \bibinfo {author} {\bibfnamefont {Y.-C.}\ \bibnamefont {Lin}},
  \bibinfo {author} {\bibfnamefont {Y.}~\bibnamefont {Yu}}, \bibinfo {author}
  {\bibfnamefont {P.}~\bibnamefont {Valencia-Acuna}}, \bibinfo {author}
  {\bibfnamefont {A.~A.}\ \bibnamefont {Puretzky}}, \bibinfo {author}
  {\bibfnamefont {R.}~\bibnamefont {Torsi}}, \bibinfo {author} {\bibfnamefont
  {C.}~\bibnamefont {Liu}}, \bibinfo {author} {\bibfnamefont {I.~N.}\
  \bibnamefont {Ivanov}}, \bibinfo {author} {\bibfnamefont {G.}~\bibnamefont
  {Duscher}}, \bibinfo {author} {\bibfnamefont {D.~B.}\ \bibnamefont
  {Geohegan}}, \emph {et~al.},\ }\href
  {https://doi.org/10.1021/acs.nanolett.0c03412} {\bibfield  {journal}
  {\bibinfo  {journal} {Nano Lett.}\ }\textbf {\bibinfo {volume} {21}},\
  \bibinfo {pages} {931} (\bibinfo {year} {2021})}\BibitemShut {NoStop}%
\bibitem [{\citenamefont {Yu}\ \emph {et~al.}(2021)\citenamefont {Yu},
  \citenamefont {Zhou}, \citenamefont {Zhang},\ and\ \citenamefont
  {Chang}}]{Chang2021}%
  \BibitemOpen
  \bibfield  {author} {\bibinfo {author} {\bibfnamefont {S.-B.}\ \bibnamefont
  {Yu}}, \bibinfo {author} {\bibfnamefont {M.}~\bibnamefont {Zhou}}, \bibinfo
  {author} {\bibfnamefont {D.}~\bibnamefont {Zhang}},\ and\ \bibinfo {author}
  {\bibfnamefont {K.}~\bibnamefont {Chang}},\ }\href
  {https://doi.org/10.1103/PhysRevB.104.075435} {\bibfield  {journal} {\bibinfo
   {journal} {Phys. Rev. B}\ }\textbf {\bibinfo {volume} {104}},\ \bibinfo
  {pages} {075435} (\bibinfo {year} {2021})}\BibitemShut {NoStop}%
\bibitem [{\citenamefont {Wang}\ \emph {et~al.}(2021)\citenamefont {Wang},
  \citenamefont {Gao},\ and\ \citenamefont {Xiao}}]{Wang2021}%
  \BibitemOpen
  \bibfield  {author} {\bibinfo {author} {\bibfnamefont {C.}~\bibnamefont
  {Wang}}, \bibinfo {author} {\bibfnamefont {Y.}~\bibnamefont {Gao}},\ and\
  \bibinfo {author} {\bibfnamefont {D.}~\bibnamefont {Xiao}},\ }\href
  {https://doi.org/10.1103/PhysRevLett.127.277201} {\bibfield  {journal}
  {\bibinfo  {journal} {Phys. Rev. Lett.}\ }\textbf {\bibinfo {volume} {127}},\
  \bibinfo {pages} {277201} (\bibinfo {year} {2021})}\BibitemShut {NoStop}%
\bibitem [{\citenamefont {Liu}\ \emph {et~al.}(2021)\citenamefont {Liu},
  \citenamefont {Zhao}, \citenamefont {Huang}, \citenamefont {Wu},
  \citenamefont {Sheng}, \citenamefont {Xiao},\ and\ \citenamefont
  {Yang}}]{Liu2021}%
  \BibitemOpen
  \bibfield  {author} {\bibinfo {author} {\bibfnamefont {H.}~\bibnamefont
  {Liu}}, \bibinfo {author} {\bibfnamefont {J.}~\bibnamefont {Zhao}}, \bibinfo
  {author} {\bibfnamefont {Y.-X.}\ \bibnamefont {Huang}}, \bibinfo {author}
  {\bibfnamefont {W.}~\bibnamefont {Wu}}, \bibinfo {author} {\bibfnamefont
  {X.-L.}\ \bibnamefont {Sheng}}, \bibinfo {author} {\bibfnamefont
  {C.}~\bibnamefont {Xiao}},\ and\ \bibinfo {author} {\bibfnamefont {S.~A.}\
  \bibnamefont {Yang}},\ }\href
  {https://doi.org/10.1103/PhysRevLett.127.277202} {\bibfield  {journal}
  {\bibinfo  {journal} {Phys. Rev. Lett.}\ }\textbf {\bibinfo {volume} {127}},\
  \bibinfo {pages} {277202} (\bibinfo {year} {2021})}\BibitemShut {NoStop}%
\bibitem [{\citenamefont {Godinho}\ \emph {et~al.}(2018)\citenamefont
  {Godinho}, \citenamefont {Reichlov{\'a}}, \citenamefont {Kriegner},
  \citenamefont {Nov{\'a}k}, \citenamefont {Olejn{\'\i}k}, \citenamefont
  {Ka{\v{s}}par}, \citenamefont {{\v{S}}ob{\'a}{\v{n}}}, \citenamefont
  {Wadley}, \citenamefont {Campion}, \citenamefont {Otxoa} \emph
  {et~al.}}]{Godinho2018}%
  \BibitemOpen
  \bibfield  {author} {\bibinfo {author} {\bibfnamefont {J.}~\bibnamefont
  {Godinho}}, \bibinfo {author} {\bibfnamefont {H.}~\bibnamefont
  {Reichlov{\'a}}}, \bibinfo {author} {\bibfnamefont {D.}~\bibnamefont
  {Kriegner}}, \bibinfo {author} {\bibfnamefont {V.}~\bibnamefont {Nov{\'a}k}},
  \bibinfo {author} {\bibfnamefont {K.}~\bibnamefont {Olejn{\'\i}k}}, \bibinfo
  {author} {\bibfnamefont {Z.}~\bibnamefont {Ka{\v{s}}par}}, \bibinfo {author}
  {\bibfnamefont {Z.}~\bibnamefont {{\v{S}}ob{\'a}{\v{n}}}}, \bibinfo {author}
  {\bibfnamefont {P.}~\bibnamefont {Wadley}}, \bibinfo {author} {\bibfnamefont
  {R.~P.}\ \bibnamefont {Campion}}, \bibinfo {author} {\bibfnamefont {R.~M.}\
  \bibnamefont {Otxoa}}, \emph {et~al.},\ }\href
  {https://doi.org/10.1038/s41467-018-07092-2} {\bibfield  {journal} {\bibinfo
  {journal} {Nature communications}\ }\textbf {\bibinfo {volume} {9}},\
  \bibinfo {pages} {1} (\bibinfo {year} {2018})}\BibitemShut {NoStop}%
\bibitem [{\citenamefont {Chen}\ \emph {et~al.}(2010)\citenamefont {Chen},
  \citenamefont {Qin}, \citenamefont {Yang}, \citenamefont {Liu}, \citenamefont
  {Guan}, \citenamefont {Qu}, \citenamefont {Zhang}, \citenamefont {Shi},
  \citenamefont {Xie}, \citenamefont {Yang}, \citenamefont {Wu}, \citenamefont
  {Li},\ and\ \citenamefont {Lu}}]{Xie2010}%
  \BibitemOpen
  \bibfield  {author} {\bibinfo {author} {\bibfnamefont {J.}~\bibnamefont
  {Chen}}, \bibinfo {author} {\bibfnamefont {H.~J.}\ \bibnamefont {Qin}},
  \bibinfo {author} {\bibfnamefont {F.}~\bibnamefont {Yang}}, \bibinfo {author}
  {\bibfnamefont {J.}~\bibnamefont {Liu}}, \bibinfo {author} {\bibfnamefont
  {T.}~\bibnamefont {Guan}}, \bibinfo {author} {\bibfnamefont {F.~M.}\
  \bibnamefont {Qu}}, \bibinfo {author} {\bibfnamefont {G.~H.}\ \bibnamefont
  {Zhang}}, \bibinfo {author} {\bibfnamefont {J.~R.}\ \bibnamefont {Shi}},
  \bibinfo {author} {\bibfnamefont {X.~C.}\ \bibnamefont {Xie}}, \bibinfo
  {author} {\bibfnamefont {C.~L.}\ \bibnamefont {Yang}}, \bibinfo {author}
  {\bibfnamefont {K.~H.}\ \bibnamefont {Wu}}, \bibinfo {author} {\bibfnamefont
  {Y.~Q.}\ \bibnamefont {Li}},\ and\ \bibinfo {author} {\bibfnamefont
  {L.}~\bibnamefont {Lu}},\ }\href
  {https://doi.org/10.1103/PhysRevLett.105.176602} {\bibfield  {journal}
  {\bibinfo  {journal} {Phys. Rev. Lett.}\ }\textbf {\bibinfo {volume} {105}},\
  \bibinfo {pages} {176602} (\bibinfo {year} {2010})}\BibitemShut {NoStop}%
\bibitem [{\citenamefont {Ma}\ \emph {et~al.}(2019{\natexlab{b}})\citenamefont
  {Ma}, \citenamefont {Xu}, \citenamefont {Shen}, \citenamefont {MacNeill},
  \citenamefont {Fatemi}, \citenamefont {Chang}, \citenamefont {Mier~Valdivia},
  \citenamefont {Wu}, \citenamefont {Du}, \citenamefont {Hsu}, \citenamefont
  {Fang}, \citenamefont {Gibson}, \citenamefont {Watanabe}, \citenamefont
  {Taniguchi}, \citenamefont {Cava}, \citenamefont {Kaxiras}, \citenamefont
  {Lu}, \citenamefont {Lin}, \citenamefont {Fu}, \citenamefont {Gedik},\ and\
  \citenamefont {Jarillo-Herrero}}]{Ma2019}%
  \BibitemOpen
  \bibfield  {author} {\bibinfo {author} {\bibfnamefont {Q.}~\bibnamefont
  {Ma}}, \bibinfo {author} {\bibfnamefont {S.-Y.}\ \bibnamefont {Xu}}, \bibinfo
  {author} {\bibfnamefont {H.}~\bibnamefont {Shen}}, \bibinfo {author}
  {\bibfnamefont {D.}~\bibnamefont {MacNeill}}, \bibinfo {author}
  {\bibfnamefont {V.}~\bibnamefont {Fatemi}}, \bibinfo {author} {\bibfnamefont
  {T.-R.}\ \bibnamefont {Chang}}, \bibinfo {author} {\bibfnamefont {A.~M.}\
  \bibnamefont {Mier~Valdivia}}, \bibinfo {author} {\bibfnamefont
  {S.}~\bibnamefont {Wu}}, \bibinfo {author} {\bibfnamefont {Z.}~\bibnamefont
  {Du}}, \bibinfo {author} {\bibfnamefont {C.-H.}\ \bibnamefont {Hsu}},
  \bibinfo {author} {\bibfnamefont {S.}~\bibnamefont {Fang}}, \bibinfo {author}
  {\bibfnamefont {Q.~D.}\ \bibnamefont {Gibson}}, \bibinfo {author}
  {\bibfnamefont {K.}~\bibnamefont {Watanabe}}, \bibinfo {author}
  {\bibfnamefont {T.}~\bibnamefont {Taniguchi}}, \bibinfo {author}
  {\bibfnamefont {R.~J.}\ \bibnamefont {Cava}}, \bibinfo {author}
  {\bibfnamefont {E.}~\bibnamefont {Kaxiras}}, \bibinfo {author} {\bibfnamefont
  {H.-Z.}\ \bibnamefont {Lu}}, \bibinfo {author} {\bibfnamefont
  {H.}~\bibnamefont {Lin}}, \bibinfo {author} {\bibfnamefont {L.}~\bibnamefont
  {Fu}}, \bibinfo {author} {\bibfnamefont {N.}~\bibnamefont {Gedik}},\ and\
  \bibinfo {author} {\bibfnamefont {P.}~\bibnamefont {Jarillo-Herrero}},\
  }\href {https://doi.org/10.1038/s41586-018-0807-6} {\bibfield  {journal}
  {\bibinfo  {journal} {Nature}\ }\textbf {\bibinfo {volume} {565}},\ \bibinfo
  {pages} {337} (\bibinfo {year} {2019}{\natexlab{b}})}\BibitemShut {NoStop}%
\end{thebibliography}%

\end{document}